\documentclass[conference,a4,10pt]{IEEEtran}

\makeatletter
\def\ps@headings{%
\def\@oddhead{\mbox{}\scriptsize\rightmark \hfil \thepage}%
\def\@evenhead{\scriptsize\thepage \hfil \leftmark\mbox{}}%
\def\@oddfoot{}%
\def\@evenfoot{}}
\makeatother 

\usepackage{graphicx}
\usepackage{subfigure}
\usepackage{amsmath}

\begin{document}

\title{Receiver Design for Realizing On-Demand WiFi Wake-up using WLAN Signals}

\author{Hiroyuki Yomo$^{\dagger}$, Yoshihisa Kondo$^{\ddag}$, Noboru Miyamoto$^{\dagger}$, Suhua Tang$^{\ddag}$, Masahito Iwai$^{*}$, and Tetsuya Ito$^{*}$\\
$^{\dagger}$Faculty of Engineering Science, Kansai University\\
%3-3-35 Yamate-cho, Suita, Osaka, 564-8680 Japan\\
%e-mail: yomo@kansai-u.ac.jp\\
$^{\ddag}$ATR Adaptive Communications Research Laboratories\\
%2-2-2 Hikari-dai, Keihan-na Science City, Kyoto, 619-0288 Japan\\
%e-mail:\{shtang, kondo\}@atr.jp\\
$^{*}$NEC Communication Systems, Ltd. NCOS Laboratory\\
%ito.tts@ncos.nec.co.jp
}

\maketitle
\begin{abstract}
In this paper, we design a simple, low--cost, and low--power
wake--up receiver which can be used for an IEEE 802.11--compliant
device to remotely wake up the other devices by utilizing its own
wireless LAN (WLAN) signals. A typical usage scenario of such a
wake--up receiver is energy management of WiFi device: a device
equipped with the wake--up receiver turns WiFi interface off when
there is no communication demand, which is powered--on only when
the wake--up receiver detects a wake--up signal transmitted by the
other WiFi device. The employed wake--up mechanism utilizes the
length of 802.11 data frame generated by a WiFi transmitter to
differentiate the information conveyed to the wake--up receiver.
The wake--up receiver is designed to reliably detect the length of
transmitted data frame only with simple envelope detection and
limited signal processing.  We develop a prototype of the wake--up
receiver and investigate the detection performance of the envelope
of 802.11 signals. Based on the obtained experimental results, we
select appropriate parameters employed by the wake--up receiver to
improve the detection performance. Our numerical results show that
the proposed wake--up receiver achieves much larger detection
range than the off--the--shelf, commercial receiver having the
similar functionality.
\end{abstract}

\section{Introduction}
Reducing the energy wastefully consumed by radio devices has
become a new challenge for wireless researchers/engineers after
the successful deployment of broadband and spectrally--efficient
radio access networks. Wireless local area network (WLAN), also
known as WiFi, is a representative example, which has shown
tremendous growth in its worldwide popularization over the last
decade as a means to provide its users with ubiquitous access to
the Internet.

One of the most common methods to reduce energy consumption of a
WiFi device is to transit WiFi interface into a sleep state during
its idle period where there is no communication demand. For
instance, a power saving (PS) mode is defined in IEEE
802.11\cite{80211}, where WiFi stations (STAs), such as laptop PC
and smartphone, transit their interfaces into a sleep mode and
periodically wake up to check demands on communications from its
associated access point (AP). However, it is difficult to adapt
the wake--up schedule to the unpredictable traffic pattern, which
inherently causes communications latency and wake--up without
actual communications demands. Therefore, the use of an extremely
low--power secondary radio has been proposed to realize
\emph{on--demand}, remote wake--up of WiFi
interface\cite{wakeonwireless}\cite{WiZi}\cite{BT1}\cite{BT2}\cite{BT3}\cite{cellular1}.
The secondary radio is in charge of wake--up signaling by which a
device sends a wake--up command to the other sleeping device. The
sleeping node turns WiFi interface on only when the wake--up
command is detected through its secondary radio. By employing a
secondary radio which consumes much smaller amount of energy than
WiFi, we can significantly reduce the amount of energy wastefully
consumed during idle periods while keeping small latency to start
communications between WiFi devices.

There have been different approaches on how to incorporate
secondary, wake--up radio into WiFi devices. Some works introduce
completely independent radio of WiFi into both sender and receiver
(e.g., ZigBee in \cite{WiZi} and Bluetooth in \cite{BT1}) while
the others exploit WiFi device at the sender side to generate
wake--up signals. A mechanism called wake--on--wlan has been
introduced in \cite{Wake-on-WLAN} where a low--power sensor mote
(802.15.4) is installed into a WiFi receiver. The sensor mote
operates at 2.4 GHz and is used to monitor the communications
activities over WLAN channels and to detect energy of WLAN
signals, which triggers the wake--up of WiFi interface. This
wake--up scheme does not require additional transmitter of
wake--up signal, however, it suffers from large probability of
false wake--up since the sensor mote uses only energy level in ISM
band to trigger the wake--up. In order to solve this problem, a
novel approach called ESENSE has been proposed in \cite{ESENSE}.
With ESENSE, 802.11 device embeds information into frame length
(length of energy burst) which is detected through energy sensing
by an 802.15.4 hardware attached to WiFi receiver. This enables
802.11 device to send specific identification (e.g., wake--up ID)
to the other sleeping device which is equipped with a secondary
802.15.4 device. We have also proposed in \cite{PIMRC_Kondo} a
mechanism for WiFi STA to send wake--up ID to a sleeping access
point (AP) which is equipped with a secondary wake--up receiver.
The proposed approach does not require each STA to install extra
hardware to generate wake--up signals while many idle APs can be
transited into sleep mode, which can reduce significant amount of
wasteful energy consumed by widely--spread WiFi
APs\cite{GreenWLAN}\cite{GLOBECOM_Tang}\cite{EURASIP_Tang}.

The communications exploiting the length of 802.11 data frames
proposed in \cite{ESENSE} and \cite{PIMRC_Kondo} require a
receiver to reliably detect the length of each transmitted frame.
In \cite{ESENSE}, the use of a commodity 802.15.4 hardware
containing CC2420 chip platform\cite{CC2420} was proposed as a
possible receiver. However, in \cite{ESENSE}, there is no
investigation on communication range achieved through energy
sensing based on CC2420--based platform. The wake--up range in
on--demand WiFi wake--up is required to be comparable to that of
WiFi data communications. If CC2420--based platform does not offer
sufficient communication range, more elaborated, yet simple
receiver is desired. On the other hand, in \cite{PIMRC_Kondo},
only simulation results were provided and there was no
investigation on receiver design and its practical feasibility.

The main contributions of this paper are twofold. First, we
investigate communication range achieved through energy sensing
with CC2420--based platform proposed in \cite{ESENSE}. With
experiments, we show that such an off--the--shelf device, which is
not specifically designed for detecting frame length, is not
sufficient to achieve wake--up range required in on--demand WiFi
wake--up. Second, based on the above observation, we design and
develop a simple, low--cost, and low--power receiver dedicated to
detecting the length of 802.11 frame. The receiver operates with a
simple envelope detection and limited signal processing. With the
developed receiver, we evaluate the basic performance for the
wake--up receiver to detect 802.11 frame length. We investigate
the impact of employed parameters on the accuracy of frame length
detection. We evaluate detection range of the designed wake--up
receiver, and show that our proposed wake--up receiver achieves
much larger detection range than CC2420-based platform and has a
potential to offer sufficient wake--up range for on--demand WiFi
wake--up.

\section{System Model and Problem Definition}
\label{sec:overview}
\subsection{Basic idea of wake-up signal transmissions}
\begin{figure}
  \centering
  \includegraphics[width=0.5\textwidth]{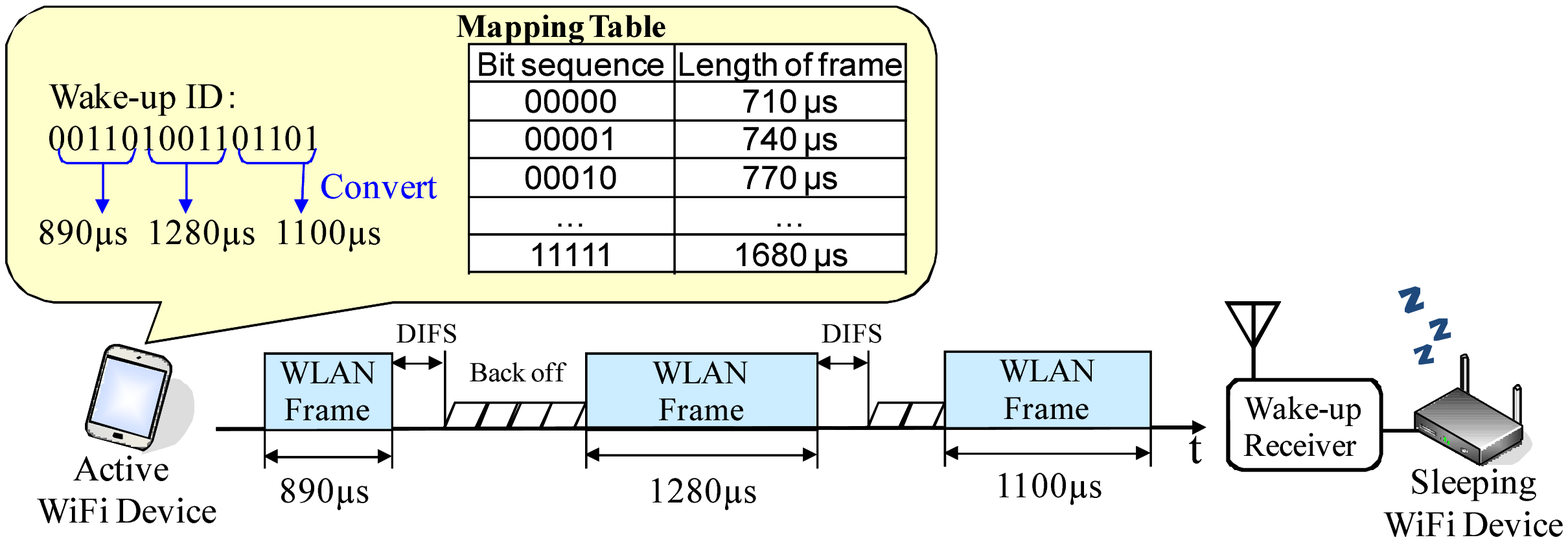}\\
  \vspace*{-0.3cm}
  \caption{System model and basic idea for conveying information through WLAN frame length.}\label{fig:system}
\end{figure}

The scenario considered in this paper is shown in
Fig.~\ref{fig:system}. Here, a WiFi device equipped with a
wake--up receiver is in a sleeping mode where WiFi interface is
completely turned off in order to save energy. The other active
WiFi device, which attempts to communicate with the sleeping
device through WiFi interface, sends a wake--up ID corresponding
to the sleeping WiFi device. Our target is to transfer information
on wake--up ID from the active WiFi device to the wake-up
receiver. The wake--up receiver should be a low--cost and
low--power device which can only employ simple
detection/demodulation scheme and is not capable of decoding
contents of WLAN data frame. The use of frame length to convey
information from WiFi device to a simple device, which has a
functionality to detect the length of energy burst, was proposed
in \cite{ESENSE}. We have also proposed a mechanism for 802.11 STA
to send information to a simple on--off--keying (OOK) receiver in
\cite{PIMRC_Kondo}. The basic idea is to embed wake--up ID into
the length of data frame transmitted by 802.11 module. We prepare
a mapping between a bit sequence and the length of WLAN frame as
shown in the example in Fig.~\ref{fig:system}. The active WiFi
device transmits frames so that the bit sequence represented by a
sequence of frames corresponds to the wake--up ID of the sleeping
device. The broadcast data are transmitted, therefore, STA does
not have to wait for the reception of ACK frames. How to avoid the
interruption by the surrounding nodes into the sequence of
wake--up frames is out of the scope of this paper (interested
readers may refer to \cite{ESENSE} and \cite{PIMRC_Kondo} for some
mechanisms to mitigate the adverse effect of such an
interruption).

\subsection{Problem Definition: Limitations of CC2420-based platform}
In order to realize information transfer using the length of
802.11 data frame, the receiver needs to detect the length of
received frame. In \cite{ESENSE}, the authors suggested using
outputs from clear channel assessment (CCA) pin of 802.15.4
receiver, which is the observed channel occupancy, in order to
detect the length of energy burst. A simple, low--cost, and
low--power platform based on CC2420 chip was proposed as a
receiver\cite{CC2420}, and the feasibility was validated through
experiments. However, there was no investigation on possible
communication range achieved by the proposed platform. Therefore,
here, we investigate the detection performance of CC2420--based
platform with different received power.

\begin{figure}
  \centering
  \includegraphics[width=0.4\textwidth]{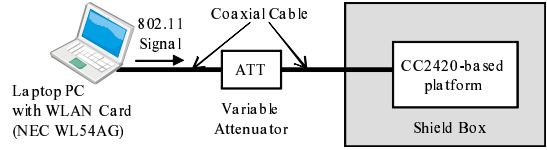}\\
  \vspace*{-0.3cm}
  \caption{Experimental setup for evaluating detection performance of CC2420-based platform.}\label{fig:CC2420_ex}
\end{figure}

Fig.~\ref{fig:CC2420_ex} shows a setup used for evaluating the
detection performance of CC2420--based platform. WLAN data frames
are generated and transmitted by a laptop PC with a WLAN card (NEC
WL54AG). The CC2420--based platform is put inside a shield box and
connected with the WLAN card using a coaxial cable. The received
signal level is controlled by adjusting a variable attenuator
attached to the coaxial cable. The transmission power of 802.11 is
fixed to be 5 dBm. Note that we use cables and shield box just to
finely tune the received signal level at the receiver. From WLAN
card, UDP packets are transmitted with IEEE 802.11b employing WLAN
data rate of 1 Mbps. We test the detection performance of two
different frame length: 800 $\mu$s (UDP payload of 12 bytes) and
1000 $\mu$s (UDP payload of 37 bytes). For each length, 10000
frames are transmitted, and we measure for each frame the number
of outputs from CCA pin of CC2420.

\begin{figure}
  \centering
  \includegraphics[width=0.5\textwidth]{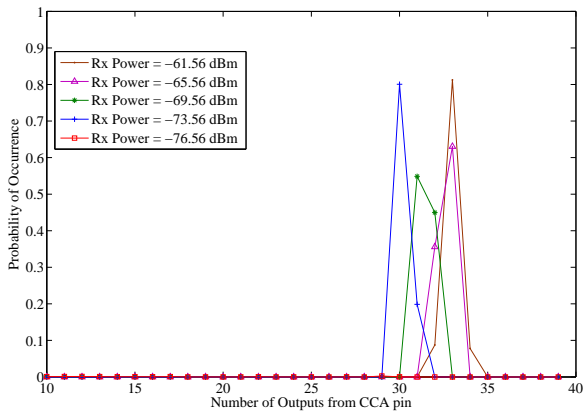}\\
  \vspace*{-0.3cm}
  \caption{Probability of occurrence of each output from CCA pin (transmitted frame length = 1000 $\mu$s).}\label{fig:1000}
\end{figure}

Figs.~\ref{fig:800} and \ref{fig:1000} show the probability of
occurrence of number of outputs from CCA pin of CC2420 with
different received power levels for 800 $\mu$s frame and 1000
$\mu$s frame, respectively. From Fig.~\ref{fig:1000}, we can see
that the number of outputs from CCA pin is 33 with the highest
probability when the received power is the largest, i.e., -61.56
dBm. The time measurement granularity of CC2420--based platform is
30.5 $\mu$s\cite{ESENSE}, therefore, 33 outputs correspond to the
measured frame length of 1006.5 $\mu$s. The other numbers of
outputs like 32 and 34 are also observed with this received power
level. However, if we allow the margin of error to be a maximum of
2 outputs, i.e., 61 $\mu$s, CC2420-based platform can reliably
identify the length of transmitted frame, i.e., 1000 $\mu$s for
this large level of received power, which is a similar result to
\cite{ESENSE}. However, looking at results with smaller received
power, we notice the following limitations of CC2420--based
platform:

\begin{figure}
  \centering
  \includegraphics[width=0.5\textwidth]{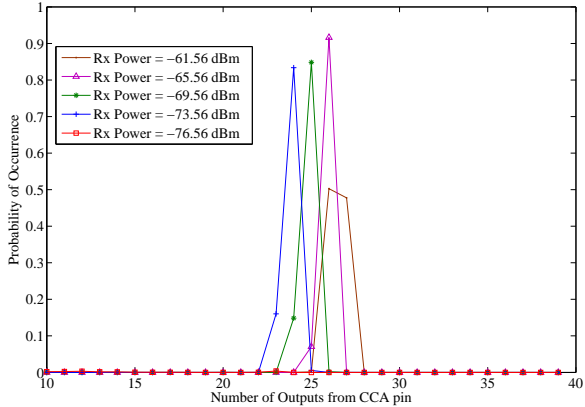}\\
  \vspace*{-0.3cm}
  \caption{Probability of occurrence of each output from CCA pin (transmitted frame length = 800 $\mu$s).}\label{fig:800}
\end{figure}

\begin{itemize}
 \item For both length, outputs from CCA pin are observed with very little
    probability for the received power level below -76.56 dBm.
    The CC2420 is designed for receiving 802.15.4 signal which has
    the bandwidth of 5 MHz while the energy of 802.11 frame is
    spread over 20 MHz. Therefore, only 25\% WLAN (5 MHz/20 MHz) signal energy
    passes the CC2420 filter. This directly reduces sensitivity level of CC2420 by 6 dBm.
    In addition, using a 5 MHz filter to receive the 20 MHz WLAN signal changes
    the envelope of WLAN frames, which degrades the performance of frame length
    detection.
    This makes it difficult for CC2420--based platform to reliably detect the frame
    length for the received power level below -76.56 dBm.
    Considering that the sensitivity level required in data communications by
    IEEE 802.11b is -90dBm@1Mbps\cite{ESENSE}, the wake--up range (range within which
    the transmission of wake-up ID is possible) achieved by
    CC2420--based platform is much smaller than data communication
    range of IEEE 802.11b. This causes an active WiFi device to fail to wake--up a
    sleeping WiFi device which can otherwise achieve WiFi
    communications with sufficiently high data rate.
    \item For both frame length, as the received power level
    becomes smaller, less number of outputs from CCA pin is observed
    with higher probability. This is due to the moving
    average employed by CC2420 for obtaining an average RSSI
    which is used to decide the output from CCA pin\cite{CC2420}.
    When the received power is small, it
    takes some period for the average RSSI to exceed
    the threshold to declare the busy channel, which results in
    less number of outputs from CCA pin. Reducing fluctuations of received signal level with moving average
    could be useful to improve the detection performance, however, it is hard to
    modify and optimize its parameter as it is implemented
    inside a chip. One way to enable the identification of each
    frame length with this limitation is to allow larger margin of errors for the
    observed outputs. For instance, if we consider that 29--35 outputs
    from CCA pin correspond to 1000 $\mu$s, the receiver can differentiate 1000
    $\mu$s frame from 800 $\mu$s frame until the received power of
    -73.56 dBm since 29 outputs are not observed when 800 $\mu$s frame is transmitted.
    However, such a large margin limits the number of
    frames used for conveying the information (the size of alphabet
    set with the terminology given in \cite{ESENSE}).
\end{itemize}

The above results show the limitations of CC2420--based platform
to be used for detecting the length of 802.11 data frame. This is
not surprising since CC2420 has been developed for data
communications following 802.15.4 standard, and the receiver
circuit and its parameters are optimized not for detecting 802.11
frame length but for supporting 802.15.4 communications under
dynamic environment even with large fluctuations of received
signal level. However, this clearly motivates us to design a
wake--up receiver dedicated to detecting 802.11 frame length,
which can achieve sufficiently large wake--up range for on--demand
WiFi wake--up.

\section{Wake--up Receiver Design for Detecting 802.11 Frame Length}
In this section, we design a wake--up receiver dedicated to
detecting the length of 802.11 data frame. The receiver should be
simple and low--cost, and operate with extremely low--power
consumption. Therefore, we employ OOK with non--coherent detection
as a basic detection scheme as often employed in wake--up receiver
designed in sensor networks\cite{WakeupR}. We add a simple
function to calculate frame length from results of detection and
signal processing to enhance the detection accuracy.

\begin{figure}
  \centering
  \includegraphics[width=0.45\textwidth]{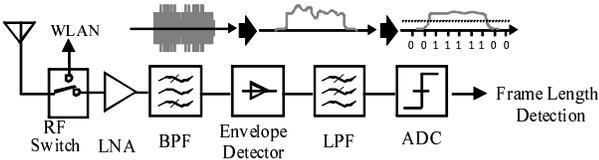}\\
  \vspace*{-0.3cm}
  \caption{A configuration of the developed wake--up receiver.}\label{fig:receiver}
\end{figure}

The block diagram of the developed wake--up receiver is shown in
Fig.~\ref{fig:receiver}. The RF switch is attached for the
wake--up receiver to share antenna with WiFi interface. With low
noise amplifier (LNA: NEC uPC8178TB, 11 dB gain) and band pass
filter (BPF: a self--developed Chebyshev filter with 20 MHz
bandwidth), the receiver passes 802.11 signals only in a specific
channel\footnote{We keep the detailed design of wake--up protocol,
including how to select a channel to transmit wake--up signals,
outside the scope of this paper.} to the envelope detector (Linear
Technology LTC5534). The samples output from the envelope detector
are smoothed with low pass filter (LPF) whose outputs are then
passed to analog to digital convertor (ADC). The impact of LPF can
be similar to moving average of CC2420--based platform, however,
here, we have room to optimize its parameter for frame length
detection, which will be discussed in detail in the following
subsection. The outputs of ADC are the results of OOK bit
detection at each sampled instance, which are used to estimate the
length of transmitted data frame. In this work, we fix the bit
detection interval to be 10 $\mu$s.

The detection of 802.11 signal is basically carried out through
the envelope detector and ADC. Each sampled value of signal
envelope is compared with a predefined threshold: if the value is
larger than the threshold, a bit "1" is detected, otherwise, "0".
While the probability to erroneously detect 1 without actual
transmissions of 802.11 signals ($p(1|0)$) depends on the noise
level and predefined threshold, the probability to miss the
transmitted signals ($p(0|1)$) is largely influenced by the
received signal strength as well as signal waveform. The signal
waveform depends on the modulation schemes employed by IEEE 802.11
standards, which are categorized into two types: single carrier
modulation and multi carrier modulation. While 802.11b adopts a
former type, which is direct sequence spread spectrum with
complementary code keying (DSSS/CCK), the other standards offering
higher rates such as 802.11a/g use the latter one, orthogonal
frequency division multiplexing (OFDM). The OFDM is known to have
large peak--to--average power ratio (PAPR) than that of single
carrier modulation~\cite{VanNee}, which means that the level of
OFDM signal fluctuates largely. In our preliminary experiment, we
have investigated the impact of signal waveform on bit detection
performance and confirmed that 802.11b signal with DSSS/CCK offers
better bit detection performance than 802.11g employing OFDM.
Therefore, in our wake--up mechanism, we utilize 802.11b for a
WiFi device to create a wake--up signal\footnote{Note that IEEE
802.11b is supported by most of the currently--available WLAN
chips to maintain backward--compatibility.}.

\subsection{Impact of LPF on bit detection performance}
In our developed wake--up receiver, in order to reduce the
fluctuation of envelope and to make the signal waveform smoother,
we introduce LPF between the envelope detector and ADC as shown in
Fig.~\ref{fig:receiver}.

\begin{figure}
  \centering
  \includegraphics[width=0.5\textwidth]{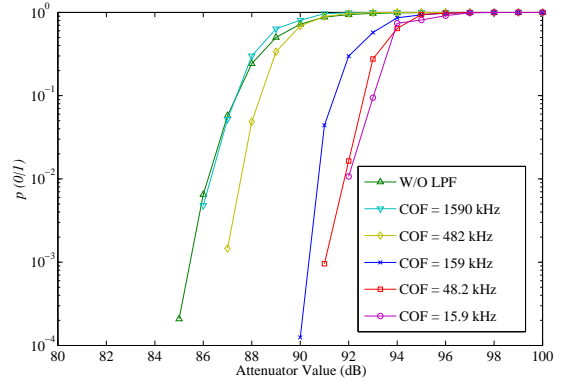}\\
  \vspace*{-0.3cm}
  \caption{The impact of cut--off--frequency (COF) of LPF on bit error probability, $p(0|1)$.}\label{fig:cutoff}
\end{figure}

As LPF, we use a very simple RC filter\footnote{More sophisticated
LPF may be used, but all the discussions given in this section can
be applied to any kind of LPF.}. Here, we investigate the impact
of cut--off--frequency (COF) of LPF on bit detection performance.
The experimental setup is similar to Fig.~\ref{fig:CC2420_ex}
except that CC2420--based platform is replaced with our developed
wake--up receiver. We vary COF of LPF by tuning the values of its
resistance and capacitance. Fig.~\ref{fig:cutoff} shows $p(0|1)$
against attenuator value (dB) for different values of COF set in
LPF. The detection threshold is adjusted so that we have
approximately $p(1|0) = 10^{-3}$ for all the attenuator values.
This figure shows a significant improvement on $p(0|1)$ as the
value of COF becomes smaller. If we compare the result employing
COF of 159 kHz with that without LPF, we have around 5 dB gain at
$p(0|1) = 10^{-3}$, and around 6 dB gain for COF of 48.2 kHz. This
gain is brought by the reduction of fluctuations within the
sampled signal. Furthermore, thanks to LPF, noise level is also
reduced and the detection threshold can be set to a lower value to
keep $p(1|0) = 10^{-3}$. This also contributes to the improvement
on $p(0|1)$ which should be decreased as the detection threshold
becomes smaller.

\subsection{Impact of LPF on the observed frame length}
\begin{figure}
  \centering
  \includegraphics[width=0.4\textwidth]{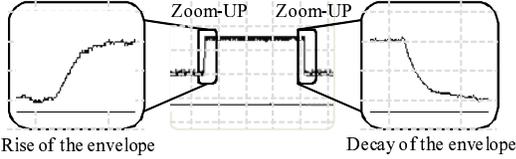}\\
  \vspace*{-0.3cm}
  \caption{A snapshot of signal waveform when LPF is applied.}\label{fig:snapshot_lpf}
\end{figure}

\begin{figure}
  \centering
  \includegraphics[width=0.5\textwidth]{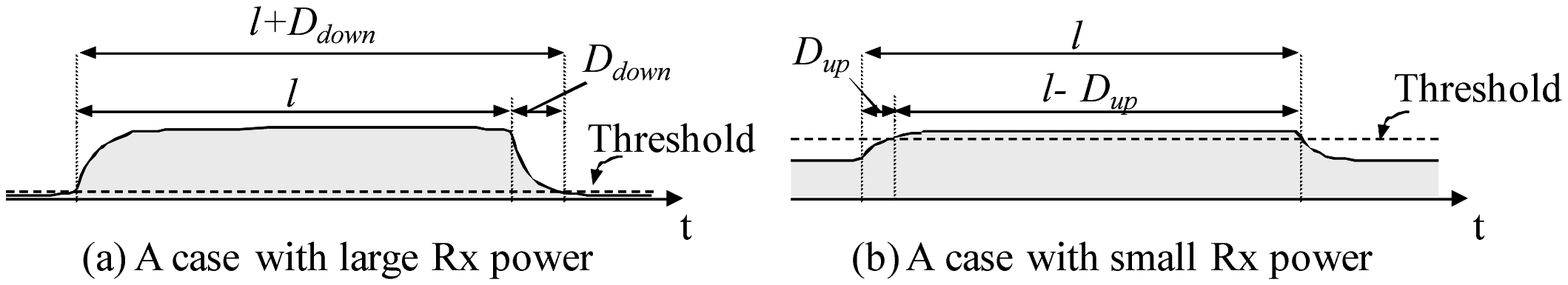}\\
  \vspace*{-0.3cm}
  \caption{The impact of LPF on the estimated frame length: (a) A case with large Rx power (b) A case with small Rx power.}\label{fig:frame_impact}
\end{figure}

Although the introduction of LPF improves the bit detection
performance, it has a side--effect that the observed frame length
becomes different from the one that is actually transmitted. This
is due to slower rise and decay caused by LPF for the head and
tail of frame envelope, respectively, as shown in
Fig.~\ref{fig:snapshot_lpf}. The frame length is estimated to be
longer than the actual one when the received power is relatively
larger than the detection threshold. An example is shown in
Fig.~\ref{fig:frame_impact} (a). Here, $l$ is the length of frame
that is actually transmitted by WiFi device. In
Fig.~\ref{fig:frame_impact} (a), while the envelope rises above
the threshold fast enough, the tail of the observed frame is
extended due to the large delay for the envelope to decay below
the detection threshold (let us define this delay as $D_{down}$).
On the other hand, when the received power is relatively small in
comparison to the detection threshold (Fig.~\ref{fig:frame_impact}
(b)), the delay for the envelope to reach the threshold ($D_{up}$)
can make the observed frame length shorter than the actual value.

Among the above problems, the extension of the observed frame
length can cause a fatal problem on the estimation on frame
length. As a wake--up signal, multiple frames may be transmitted
sequentially as shown in Fig.~\ref{fig:system}. In order to
reliably estimate a frame length, inter--frame space (i.e., at
least one "0" between two succeeding frames) must be detected
besides the correct detections of all bits of "1" constituting a
single frame. The shortest inter--frame space can be observed when
WiFi device picks up a back--off counter of 0, which results in
DIFS between two succeeding frames. If $D_{down}$ caused by the
introduction of LPF is large enough to mask DIFS, it becomes
impossible for the wake--up receiver to detect a space between two
succeeding frames. In fact, considering that WiFi device and the
wake--up receiver are not synchronized, we have to keep space of
at least sampling interval, which is 10 $\mu$s in this study,
between succeeding frames. Since DIFS is 50 $\mu$s, $D_{down}$
must be less than 40 $\mu$s. Table~\ref{table:COF_Ddown} shows
$D_{down}$ for different COF measured by our prototype. For each
value of COF, we conduct 10 measurements, and show minimum,
maximum, and average values of $D_{down}$. The received power is
set to be -10.2 dBm which is almost the same value as the maximum
received signal power assumed in IEEE 802.11
standard~\cite{80211}, i.e., -10 dBm\footnote{This value is
extremely large. Considering the transmission power of 802.11
module and well--known propagation model like two--ray path loss
model, the distance between transmitter and receiver to have such
a large received power is far less than one meter, which in fact
does not require remote wake--up of WiFi device.}. From this
table, we can see that smaller values of COF make $D_{down}$
larger, and COF of 15.9 kHz and 48.2 kHz have $D_{down}$ larger
than 40 $\mu$s for all the minimum, maximum, and average values.
Therefore, these values of COF are not applicable though they have
better bit detection performance. On the other hand, the average
values for COF of 482 kHz and 1590 kHz are less than 40 $\mu$s,
however, bit error probabilities for these COF are high as seen in
Fig.~\ref{fig:cutoff}. The COF of 159 kHz has the maximum and
average $D_{down}$ larger than 40 $\mu$s, however, its average
value is close to 40 $\mu$s. Furthermore, considering that random
back--off with contention window (CW) is applied, the minimum
separation of 40 $\mu$s between succeeding frames occurs with low
probability. Therefore, COF of 159 kHz can be a good candidate
considering the trade--off between bit error performance and space
detection, and is used for evaluating the detection performance of
the developed wake--up receiver in the next subsection.

\begin{table}[t]
\begin{center}
\caption{Measurement results of minimum, maximum, and average
$D_{down}$ for different COFs.}
\begin{tabular}{|c|c|c|c|} \hline
COF & minimum ($\mu$s) & maximum ($\mu$s) & average ($\mu$s)
\\ \hline
15.9 kHz & 149.6 & 192.2 & 168.98 \\ \hline 48.2 kHz & 70.2 &83.4
& 77.5 \\ \hline 159 kHz & 35.8 & 57.4 & 47.24 \\ \hline 482 kHz &
4.2 &85.8 & 12.76 \\ \hline 1590 kHz & 3 &3.8 & 3.3 \\ \hline
\end{tabular}
\label{table:COF_Ddown}
\end{center}
\end{table}

\subsection{Detection Performance of developed wake--up receiver}
\label{sec:design} 
\begin{figure}
  \centering
  \includegraphics[width=0.25\textwidth]{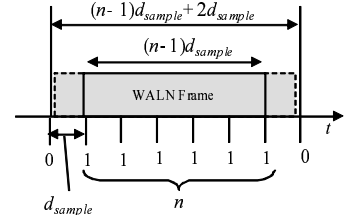}\\
    \vspace*{-0.3cm}
  \caption{The impact of asynchronous bit detection on estimated frame length.}\label{fig:sample}
\end{figure}

Here, we investigate detection range of the
developed wake--up receiver. We examine frame length detection
error rate (probability that the frame length is not detected
correctly) for three different frame length, 720 $\mu$s, 800
$\mu$s, and 1000 $\mu$s. We allow the margin of error of $\pm$30
$\mu$s for frame length detection. For instance, for 720 $\mu$s,
if the continuous detection of "1" is observed for 69--75 times,
we consider that 720 $\mu$s frame is transmitted by WLAN card
(Recall that the bit detection interval of the developed wake--up
receiver is 10 $\mu$s). Note that this resolution is the same as
the one used in \cite{ESENSE}, therefore, we can define the same
alphabet size considered in \cite{ESENSE}. This margin is used for
accommodating the impact of LPF on the observed frame length as
discussed in the previous subsection. Furthermore, since WiFi
device sending a wake--up signal and wake--up receiver are not
synchronized with each other, there can be a maximum error of $2
\times d_{sample}$ if the frame length is estimated from the
number of succeeding detections of "1", where $d_{sample}$ is the
sampling interval (see Fig.~\ref{fig:sample}). The error margin is
used to alleviate the adverse effect of such an asynchronous
transmission.

\begin{figure}
  \centering
  \includegraphics[width=0.5\textwidth]{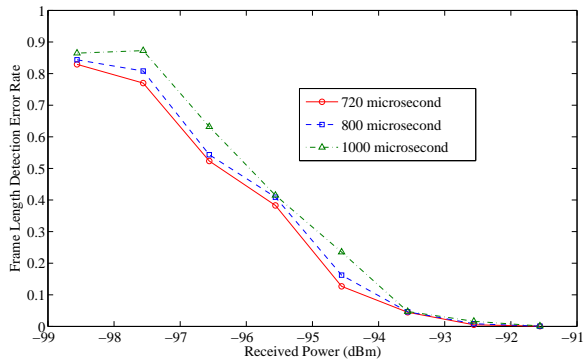}\\
  \vspace*{-0.3cm}
  \caption{Frame Length Detection Error Rate against Received Power for the developed wake--up receiver.}\label{fig:frame_length_error}
\end{figure}

Fig.~\ref{fig:frame_length_error} shows the frame length detection
error rate against received signal power for three different frame
length. In this experiment, 10000 frames of each length are
transmitted. From this figure, we can first see that the detection
error rate is lower for shorter frame length. This is because we
need more number of correct detections of "1" for correctly
detecting longer frame. The detection error rate is deteriorated
as the received power becomes smaller, however, the figure shows
that the correct detection of frame length is possible with high
probability even with the received power below -90 dBm. This means
that within data communication range of 802.11b (sensitivity level
of -90 dBm), our developed wake--up receiver can reliably detect
the length of 802.11 frame transmitted by the active WiFi device.
Therefore, successful wake--up of sleeping WiFi device is possible
with high probability whenever data communications with
sufficiently high data rate are possible. Thus, our developed
wake--up receiver can meet the requirement to be employed for
on--demand WiFi wake--up.

\subsection{Discussions on power consumption}
We have also measured power consumption of our developed wake--up
receiver and found out that its power consumption is approximately
30 mW. Considering that CC2420--based platform has the power
consumption of 60 mW\cite{ESENSE}, we can say that our wake--up
receiver operates with low power consumption. However, its value
is still higher than the other wake--up receivers developed in the
research field of sensor network, which operates less than 1 mW.
Note that these wake--up receivers for sensor network have the
optimized circuit configuration to reduce their power consumption.
Our developed wake--up receiver is still a prototype and has much
room to reduce its power consumption by optimizing circuit
configuration and choosing appropriate components, which is kept
for our future work.

\section{Conclusions}
\label{sec:conclusions} In this paper, we have designed a simple,
low--cost, and low--power wake--up receiver dedicated to detecting
802.11 frame length. This type of receiver can be applied to
reduce wasteful energy consumed by WiFi devices without installing
specialized hardware to transmit wake--up signals. We have
experimentally investigated the detection performance of the
developed receiver which is capable of making only simple envelope
detection and limited signal processing. We have tuned parameters
of the developed wake--up receiver based on the measurement
results. Our numerical results have shown that our proposed
wake--up receiver can achieve larger detection range than the
commodity CC2420 receiver which has functionality to detect the
length of energy burst and previously proposed as a receiver in
the similar setting.

Our future work includes the investigation of detection
performance in a practical wireless environment, and the design of
wake--up protocols to validate the system--level feasibility of
our wake--up approach.

\section*{Acknowledgement} This work is supported by
the Strategic Information and Communications R\&D Promotion
Programme (SCOPE) funded by Ministry of Internal Affairs and
Communication, Japan.

\bibliographystyle{IEEEtran}
\bibliography{IEEEabrv,rod}

\end{document}